\documentclass[refree]{aa}
\usepackage{graphics}
\begin{document}
\thesaurus{06(02.16.2; 08.03.4; 13.09.6)}
\title{Polarization measurements of Vega-like stars}

\author{H. C. Bhatt and P. Manoj}
\institute{Indian Institute of Astrophysics, Bangalore 560 034, India}
\offprints{H. C. Bhatt}
\mail{hcbhatt@iiap.ernet.in, \\
 manoj@iiap.ernet.in }
\date{}
\maketitle
\begin{abstract}
Optical linear polarization measurements are presented for
about 30 Vega-like stars. These are then compared with the polarization 
observed for normal field stars. A significant fraction of the Vega-like
stars are found to show polarization much in excess of that expected to 
be due to interstellar matter along the line of sight to the star. The
excess polarization must be intrinsic to the star, caused by
circumstellar scattering material that is distributed in a flattened
disk. A correlation between infrared excess and optical polarization is 
found for the Vega-like stars.
\end{abstract}
\section{Introduction}
The Infrared Astronomy Satellite (IRAS) detected (Aumann et al. 1984) a
large infrared excess at wavelengths longward of $12\mu $ from the otherwise 
normal main-sequence A0V star Vega ($\alpha $ Lyr). The infrared excess is
thought to be due to emission from circumstellar dust, in thermal
equilibrium with the stellar radiation from the central star, and
distributed in a shell or ring, several tens to hundred AU in
extent. Since then a large number of main-sequence stars have been found
to have similar infrared excesses in the IRAS wavebands and are called
Vega-like stars (see eg. Backman \& Paresce 1993, Vidal-Madjar \& Ferlet
1994, Lagrange-Henri 1995, Mannings \& Barlow 1998). \\

\begin{table*}
\caption[]{Polarization measurements of Vega-like stars}
\vspace{0.25cm}
\begin{tabular}{rlllrlllllrr}\hline \hline
HD  &    Date of &     Spectral &   V &     B-V &     E(B-V) &     m-M &   Ref &    P(\%) &    $\epsilon_P$(\%)
&   $\theta (^{\circ})$ & $\epsilon_{\theta}(^{\circ})$  \\  number &  Observation &   Type &     mag & mag  & mag & & & & & & \\          
&  (1999)& & & & & & & & & & \\ \hline \hline

9672 &     17Jan &      A1V  &     5.6 & 0.06  &   0.04  &   3.94 &   H   &    0.17 &  0.05 &  12  &  8 \\
17206 &    17Jan &      F5/F6V &   4.4 & 0.48   &  0.03  &   0.73 &   H   &    0.10 &  0.03 &  88  &  5 \\
17443 &    17Jan &      B9V &      8.7 & 0.30  &   0.38  &   7.67 &   H   &    1.10 &  0.17 &  139 &  5 \\
32509 &    16Jan &      A2 &       7.5 & 0.20   &  0.15  &   5.89 &   H   &    0.95 &  0.09 &  51  &  3  \\
34700 &    17Jan &      G0 &       9.1 &  0.56   &  0.00 &   10.32 &  H   &    0.11 &  0.12 &  37  & 19  \\
37389 &    17Jan &      A0  &      8.3 & -0.06   & 0.00 &    7.04 &      &     0.46&   0.12&   15 &  8   \\
53300 &    16Jan &      A0  &      8.0 &  0.31    & 0.31 &    5.79 &      &     1.00&   0.13&   81 &   3 \\
93331 &    08Apr &      B9.5V &    7.2 &  0.02  &   0.07 &    5.90 &   H  &     0.24&   0.06&   175&   5 \\
98800  &   13Mar &      K5  &      8.8 &  1.15   &  0.10 &    3.34 &   H  &     0.54&   0.13&   89 &   6 \\
99211 &    08Apr &      A9V  &     4.0 &  0.21    & 0.00 &    2.05 &   H  &     0.06&   0.06&   175&   5 \\
102647 &   07Apr &      A3V &      2.1 &  0.09   &  0.01 &    0.22 &   H  &     0.15&   0.05&   50 &   6 \\
109085 &   08May &      F2V  &     4.3 &  0.38   &  0.03 &    1.30 &   H  &     0.38&   0.04&   147&   5 \\
115892 &   08Apr &      A2V &      2.7 &  0.06   &  0.18 &    1.27 &   H  &     0.18&   0.06&   164&   6 \\ 
121847  &  08May &      B8V &      5.2 & -0.08   &  0.01 &    5.09 &   H  &     0.53&   0.04&   172&   2 \\
123247  &  08May &      B9V &      6.4 &  0.00   &  0.07 &    5.02 &   H  &     0.21&   0.06&   57 &   5 \\
131885  &  08Apr &      A0V &      6.9 &  0.01   &  0.01 &    5.42 &   H  &     0.57&   0.07&   13 &   3 \\
135344  &  11Mar &      A0V &      8.6 &  0.48   &  0.48  &    5.82 &      &     0.07&   0.11&   108&   15 \\
139614  &  10Mar  &     A7V  &     8.2 &  0.23   &  0.03 &    5.76 &      &     0.65&   0.10&   38 &   5  \\
139664  &  12Mar &      F5IV/V &   4.6 &  0.41  &   0.00  &    1.22 &   H  &     0.76&   0.07&   165&   5  \\
142096  &  06may &      B3V &      5.0 & -0.01   &  0.18 &    5.19 &   H  &     0.42&   0.04&   173&   2   \\
142114  &  08May &      B2.5V &    4.5 & -0.07  &   0.15 &    5.62 &   H  &     0.45&   0.05&   14 &   2 \\
142165  &  06May &      B5V &      5.3 & -0.01   &  0.15 &    5.52 &   H  &     0.75&   0.06&   3  &   2 \\
142666  &  11Mar &      A8V &      8.8 &  0.52   &  0.27 &    5.51 &      &     0.80&   0.12&   72 &   4 \\
143006  &  08Apr &      G6/G8 &    10.2 &  0.73 &   0.03 &    4.82 &      &     0.69&   0.08&   7  &   3 \\
145482  &  08may &      B2V &      4.5 & -0.17    & 0.06 &    5.78 &   H  &     0.50&   0.05&   117&   4 \\
149914 &   11Mar &      B9.5IV &   6.7 &  0.24    & 0.29 &    6.09 &   H  &     2.54&   0.07&   66 &   1 \\ 
233517  &  10Mar &      K2  &      9.7  & 1.30    & 0.40 &    2.16 &      &     1.80&   0.21&   177&   4 \\ \hline

\end{tabular} \\

\noindent H in the Ref. column refers to the{\it Hipparcos}
catalogue
\end{table*}

\begin{table*}
\caption[]{Polarization data for additional Vega-like stars from literature}
\vspace{0.25cm}
\begin{tabular}{rllrrlllr}\hline \hline

     HD &  P(\%) & $\epsilon_P$(\%)  & $\theta (^{\circ})$    & $\epsilon_{\theta}(^{\circ})$   &  E(B-V) &  V & Spectral &  m-M \\
   Number &  &  &    &    &    mag & mag & Type &   \\  \hline \hline                 

    3003 &  0.028 &    0.009 &  127.7  &   9.1 &   0.00   &  5.2   &   A2   &    2.31  \\
   10476 &  0.050 &    0.035 &  178.1  &  19.3 &   0.00  &   5.2   &   K1V  &  -0.59   \\
   10647 &  0.040 &    0.035 &   20.7  &  23.6 &   0.00  &   5.6   &   G0I  &   10.20  \\
   10700 &  0.005 &    0.007 &  148.9  &  35.0 &   0.00  &   3.6   &   G8V  &  -2.47    \\
   14055 &  0.030 &    0.120 &   18.0  &  63.4 &   0.00  &   4.0   &   A0V  &   2.84  \\
   16157 &  0.131 &    0.047 &  170.6  &  10.2 &   0.00  &   8.6   &   K7V  &   0.39   \\ 
   18978 &  0.006 &    0.006 &   10.8  &  26.6 &   0.00  &   4.2   &   A3V  &   2.07    \\
   20010 &  0.006 &    0.009 &  115.3  &  36.9 &   0.00  &   4.0   &   F8IV &   1.15   \\
   22049 &  0.007 &    0.010 &  147.6  &  35.5 &   0.00  &   3.7   &   K2V  &  -2.47   \\
   38393 &  0.570 &    0.035 &   42.5  &   1.8 &   0.00  &   3.6   &   F6V  &   0.00   \\
   43955 &  0.090 &    0.035 &   63.1  &  11.0 &   0.10  &   5.3   &   B2V  &   7.59    \\
   49336 &  1.300 &    0.035 &    2.0  &   0.8 &   0.10  &   6.2   &   B3V  &   7.66    \\
   53376 &  0.014 &    0.037 &  175.5  &  54.0 &   0.01  &   9.3   &   F3/F5V & 6.10    \\
   68456 &  0.060 &    0.035 &  166.4  &  16.3 &   0.00  &   4.8   &   F7V  &   1.50    \\
   69830 &  0.080 &    0.035 &  140.6  &  12.3 &   0.00  &   6.0   &   G8V  &   0.60    \\
   71155 &  0.020 &    0.120 &   13.0  &  71.6 &   0.00  &   3.9   &   A0V  &   2.95    \\ 
   73390 &  0.170 &    0.035 &  119.5  &   5.9 &   0.10  &   5.3   &   B3V  &   6.73    \\
   95418 &  0.000 &    0.120 &   43.0  &  90.0 &   0.00  &   2.4   &   A1V  &   1.27    \\
  108483 &  0.140 &    0.035 &   59.5  &   7.1 &   0.00  &   3.9   &   B2V  &   5.99     \\
  114981 &  0.140 &    0.035 &   32.3  &   7.1 &   0.10  &   7.1   &   B7   &   7.73     \\
  128167 &  0.090 &    0.120 &   26.0  &  33.7 &   0.00  &   4.5   &   F2   &   1.71     \\
  139006 &  0.060 &    0.120 &   29.0  &  45.0 &   0.00  &   2.2   &   A0V  &   1.27    \\
  143018 &  0.348 &    0.107 &   37.2  &   8.7 &   0.10  &   2.9   &   B1V  &   6.26    \\
  149630 &  0.080 &    0.120 &   75.0  &  36.9 &   0.10  &   4.3   &   B9V  &   3.16    \\
  153968 &  0.901 &    0.141 &   33.2  &   4.5 &   0.08  &   9.3   &   F0V  &   7.70    \\ 
  161868 &  0.008 &    0.001 &   33.3  &   3.6 &   0.10  &   3.8   &   A0V  &   2.38   \\
  162917 &  0.100 &    0.035 &   65.2  &   9.9 &   0.00  &   5.7   &   F4V  &   1.99    \\
  172167 &  0.020 &    0.120 &   75.0  &  71.6 &   0.00  &   0.0   &   A0V  &  -1.10    \\
  176638 &  0.030 &    0.035 &   13.5  &  30.3 &   0.00  &   4.7   &   A0V  &   3.99    \\
  181296 &  0.040 &    0.035 &   99.9  &  23.6 &   0.00  &   5.0   &   A0V  &   3.90   \\
  181869 &  0.010 &    0.035 &   75.9  &  60.3 &   0.00  &   4.0   &   B9II &   4.40   \\
  214953 &  0.025 &    0.015 &  115.3  &  16.7 &   0.00  &   6.3   &   G0V  &   1.15   \\
  216956 &  0.005 &    0.005 &  122.9  &  26.6 &   0.00  &   1.3   &   A3V  &  -0.77   \\
  224392 &  0.041 &    0.011 &  128.5  &   7.6 &   0.00  &   5.2   &   A2V  &   2.65    \\  \hline

\end{tabular}
\end{table*}

\begin{figure*}
\resizebox{\hsize}{!}{\includegraphics{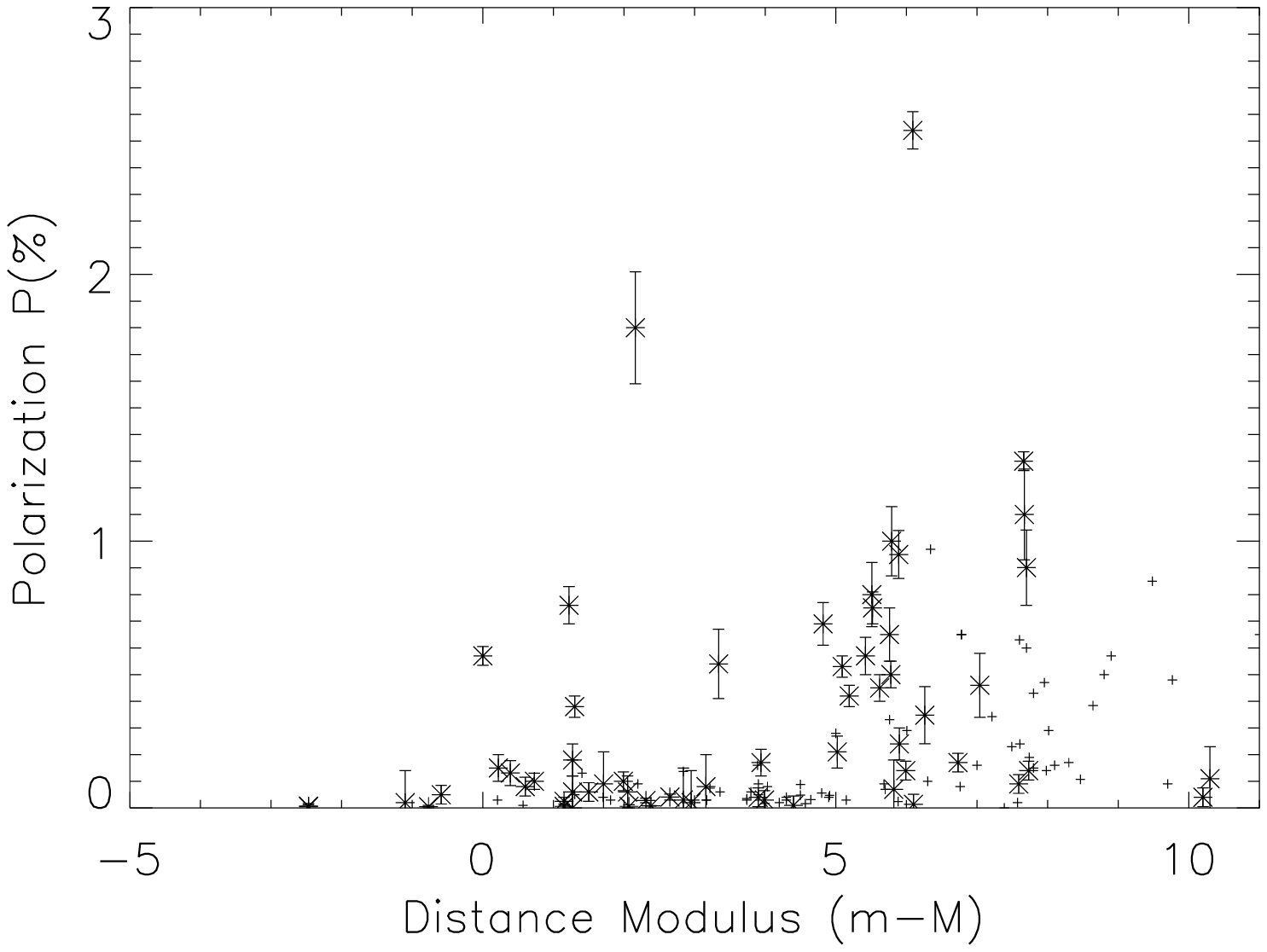}}
\caption{Percentage polarization plotted against distance modulus for Vega-like stars. 
Vega-like stars are represented by the asterix symbol while crosses are  used for normal stars}
\label{Figure 1.}
\end{figure*}   

What is the spatial distribution of the circumstellar dust in Vega-like
stars? Optical coronagraphic observations of $\beta$ Pic  by Smith \&
Terrile (1984) showed that the scattering dust in this object is
distributed in a flattened disk being viewed nearly edge-on. Vega-like
stars are now generally thought to have circumstellar dust distributed in 
disks. The disk structures in these objects may be the end products of
evolution of more massive disks associated with pre-main-sequence stars
and are replenished by the dust debris produced by the disruption of
planetesimals due to collisional and thermal evaporation processes (eg.
Backman \& Paresce 1993, Malfait et al. 1998). However, other than
for $\beta $ Pic, direct evidence for the flattened disk-like distribution 
of the circumstellar material around Vega-like stars has so far been obtained
only for a handful of these objects by optical and infrared imaging ( BD
+31$^{\circ}$643, Kalas \& Jewitt 1997; SAO 26804, Skinner et al. 1995;
HR 4796A, Jayawardhana et al. 1998 and Koerner et al. 1998) \\

\begin{figure*}
\resizebox{\hsize}{!}{\includegraphics{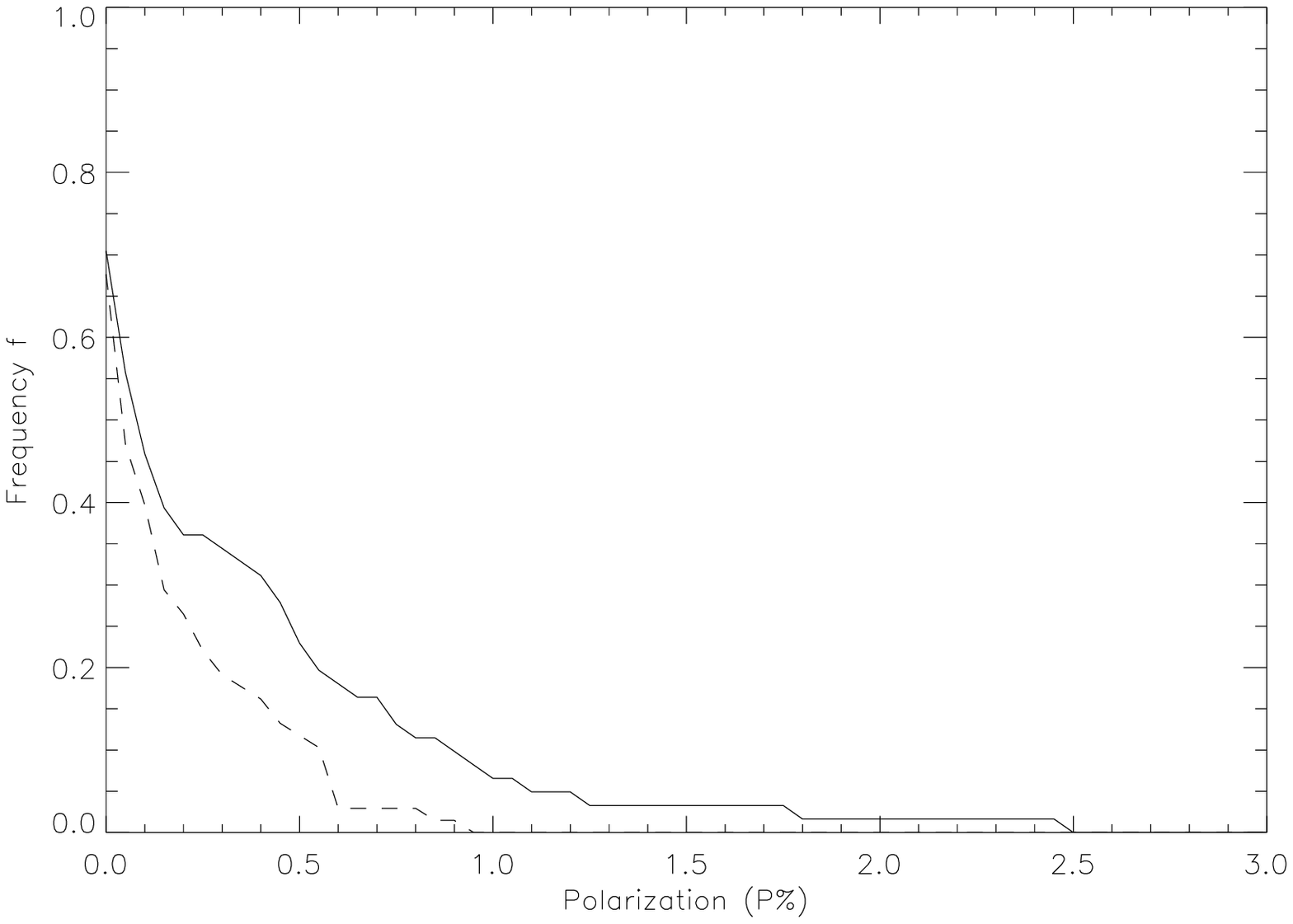}}
\caption{Frequency distribution of observed polarization values
for Vega-like and normal field stars. $f$ is the fractional number
 of stars with observed polarization larger than a given value. 
Solid line represents Vega-like stars and dashed line normal field stars}
\label{Figure 2.}
\end{figure*}   
 
Circumstellar dust emits thermally in the infrared producing the infrared
excess. It also scatters the stellar radiation giving rise to reflection
nebulosity. Another manifestation of scattering by dust in the
circumstellar disk can be polarization of the stellar radiation. For
example, the polarization observed in the light of young T Tauri stars and
the Herbig Ae/Be stars is generally ascribed to dusty circumstellar disks
(e.g. Bastien 1988). Therefore, it is of interest to look for polarization
in Vega-like stars. In $\beta $ Pic, where the disk can be resolved,
imaging polarimetry in the $R$ band shows linear polarization at the level 
of $\sim 17 \% $ (Gledhill et al. 1991). When the disk
can not be resolved, any polarization in the integrated light from the
{\it star + disk} system will show much lower values of polarization
depending on the amount of scattering dust, degree of flattening of the
disk and its orientation with respect to the observer's line of sight to
the star. In the observed polarization for an object, there will also be
present a component of the interstellar polarization that will depend on
the direction and distance to the star. Any significant intrinsic
component of polarization in the observed polarization for a star will 
indicate the presence of circumstellar dust with a spatial distribution
around the star that is not spherically symmetric. The dust could be 
in a disk-like structure with the disk-plane making relatively small
angles with the line of sight, because a circularly symmetric disk at right
angles to the line of sight will produce no net polarization in the
integrated light. Thus polarization measurements can give important 
information on the spatial distribution of scattering dust in Vega-like
stars. \\

 \begin{figure*}
\resizebox{\hsize}{!}{\includegraphics{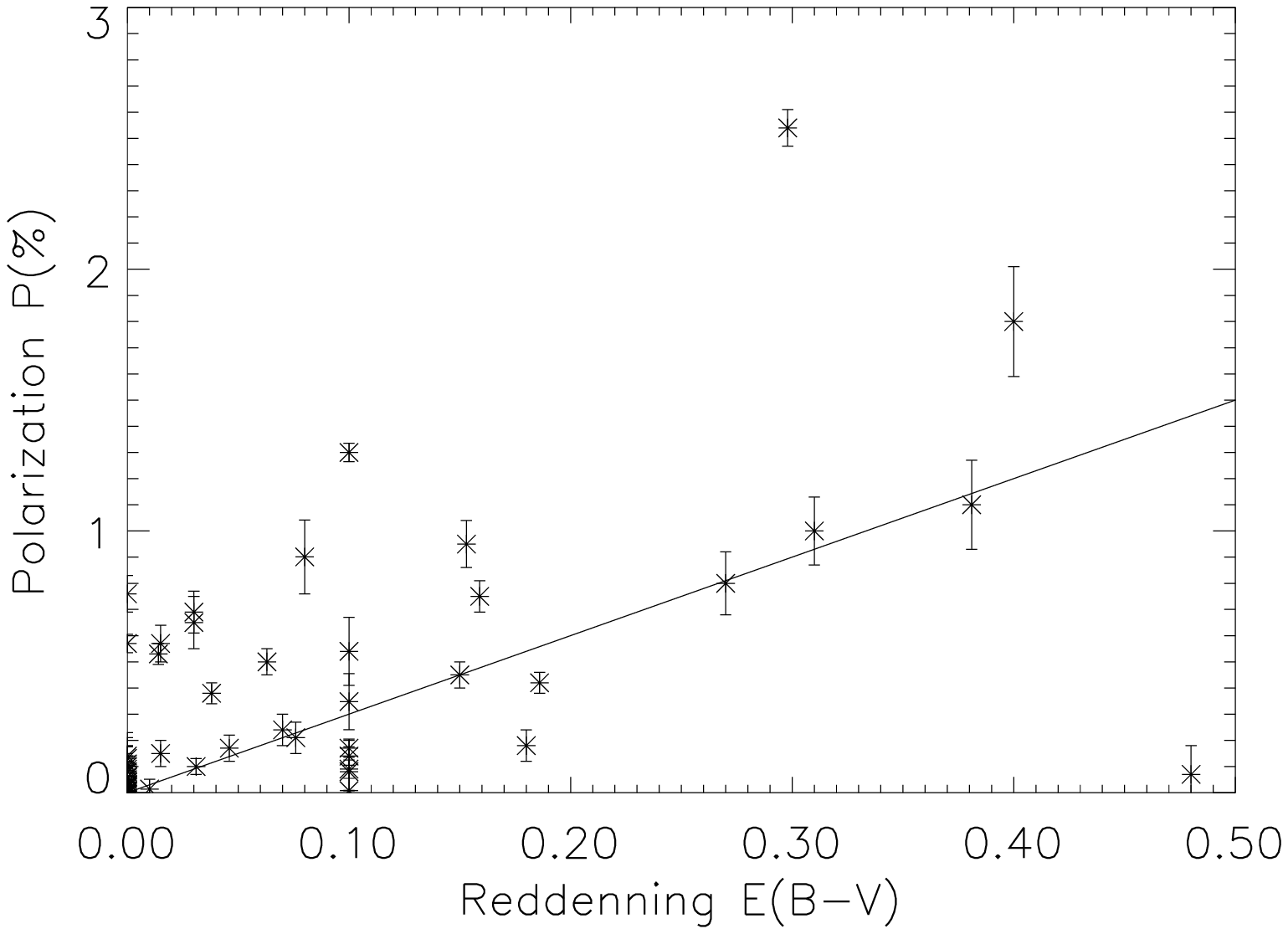}}
\caption{Percentage polarization  against reddenning 
for Vega-like stars.The average interstellar relation $ P(\%)/E(B-V)=3$ is shown as the solid line}
\label{Figure 3.}
\end{figure*}

In this paper we present the results of optical linear polarization
measurements of about 30 Vega-like stars. We also compile polarization 
data on additional Vega-like stars from the literature. The Vega-like
stars are then compared with normal field stars. It is found that a
significant fraction of the Vega-like stars show polarization that is much
larger than can be explained as due to the interstellar polarization.
In {\it Section 2} we present our measurements. Comparison with field
stars, the intinsic polarization of Vega-like stars, the distribution 
and nature of the dust grains is discussed in {\it Section 3}. The
conclusions are summarized in {\it Section 4}.

\section{Observations}
Optical linear polarization measurements were made with a fast
star-and-sky chopping polarimeter (Jain \& Srinivasulu 1991) coupled at
the $f/13$ Cassegrain focus of the 1-m telescope at the Vainu Bappu
Observatory, Kavalur of the Indian Institute of Astrophysics. A dry-ice
cooled R943-02 Hamamatsu photomultiplier tube was used as the detector.
All measurements were made in the $V$ band with an aperture of 15 {\it arcsec}. 
Observations were made during the period of January to May 1999. The instrumental
polarization was determined by observing unpolarized standard stars  
from Serkowski (1974). It was found to be $\sim 0.1\%$, and has been
subtracted vectorially from the observed polarization of the programme
stars. The zero of the polarization position angle was determined by
observing the polarized standard stars from Hsu \& Breger (1982). The
position angle is measured from the celestial north, increasing eastward.   
The Vega-like stars selected for observations were taken from the lists
of Vega-like objects in Backman \& Paresce (1993), Coulson et al. (1996) and
Mannings \& Barlow (1998). \\

The results of our polarization measurements are presented in Table 1.
We observed  27 Vega-like stars. In Table 1, the HD numbers of the stars
observed are given in Column 1, the date of observation in Column 2,
spectral type in Column 3, $V$ magnitude in Column 4, the colour $B - V$
in Column 5, colour excess $E(B - V)$ in Column 6 and the distance modulus
$m - M$ in Column 7. For most of the stars the stellar parameters and the
the distance modulus are taken from the $Hipparcos$ catalogue (ESA 1997)
as indicated by $H$ in the reference Column 8. Stars for which $Hipparcos$
distances are not available, the distance modulus is estimated from 
spectral type and photometric magnitudes. Intrinsic colours and absolute
magnitudes are taken from Schmidt-Kaler (1965). The distance modulus has
been corrected for extinction by using a value of $3.0$ for the ratio of
total to selective extinction $A_{V}/E(B - V)$. The extinction values are
generally quite small. Negative values of $E(B -V)$ in Column 6 have been 
called zero to derive the corrected magnitudes of the stars. The observed
degree of polarization $ P (\%)$ and the position angle $\theta
(^{\circ})$ are given in Columns 9 and 11, while the probable errors in
the measurements of polarization $\epsilon_{P}$ and the position angle
$\epsilon_{\theta}$ are given in Columns 10 and 12.

\section{Discussion}
It can be seen from Table 1 that the degree of polarization for the
Vega-like stars varies from  small values of $\sim 0.06 \%$ (for HD 99211)
to values as large as $\sim 2.5 \%$ (for HD 149914). Most of the Vega-like 
stars are nearby objects within $\sim $ 100 {\it pc} and are at large galactic
latitudes ($|b|$ generally $> 10^{\circ}$, due to a selection effect in
making the searches for stars with excess far-infrared fluxes in the
IRAS catalogue). At such short distances the contribution of the interstellar 
polarization to the observed polarization is expected to be small $\sim
0.1 \%$. A significant fraction (about 2/3) of the Vega-like stars have
polarization values $>0.3 \%$. As discussed below, such large 
values indicate intrinsic polarization being caused by the circumstellar
dust around the Vega-like stars.

As most of the known Vega-like stars are relatively bright objects, a
number of them are likely to have been observed earlier as part of other  
polarimetric programmes. Therefore, to supplement our observations, we
have made a search for polarization measurements of additional Vega-like
stars in the $SIMBAD$ data base at $CDS$, Strasbourg. We found polarization
measurements for 34 such stars in the catalogue of stellar polarization by Heiles (2000).
In a manner similar to Table 1, we list data for these stars from Heiles (2000) in Table 2.
 Altogether, in Tables 1 and 2, we
now have 61 Vega-like stars with polarization measurements. In the
following we compare the polarimetric behaviour of the Vega-like stars
with normal field stars.

In order to assess the strength of the intrinsic component in the
polarization of the Vega-like stars, one needs to have an estimate of the
interstellar polarization. For more distant Vega-like stars the
interstellar component could be relatively large, since the interstellar
polarization in general increases with distance, and should be subtracted
from the observed polarization. This is , however, not possible for the
programme stars individually, as the interstellar polarization in the
direction of these stars, as a function of distance, is generally not
known. We therefore make a statistical comparison between the polarization
observed for the Vega-like stars and normal field stars with similar
magnitudes and distances. For the comparison, polarization
measurements for normal field stars have been taken from 
Heiles (2000). Normal stars within $\sim 1^{\circ}$ of the
Vega-like stars were chosen for comparison. Stars with any known
peculiarities (like the presence of emission lines in the spectra, infrared 
excesses, association with nebulosities,variability etc.) were excluded.
In Fig. 1 we plot the observed degree of polarization against the distance 
modulus for the Vega-like stars as well as the normal field stars. It 
can be seen from Fig. 1 that, on an average, the Vega-like stars have
polarization values generally larger than the normal field stars at
comparable distances.
               The average value of polarization for the
Vega-like stars is 0.34\% with a large dispersion of 0.48\%,
while the normal field stars have an average polarization of
 0.2\% and a smaller dispersion of 0.22 \%. Fig. 2 shows 
frequency distribution (as a fraction of the total number of
stars in the two samples seperately) of stars with observed
polarization larger than a given value. Now the difference
between the polarimetric behaviour of the two samples of stars
is more clear. About 30\% of the Vega-like stars show
polarization values larger than 0.5\%, whereas only 13\% of the
normal field stars show polarization values larger than 0.5\%.
There are no field stars with polarization values larger than
1\%, while about 10\% of the Vega-like stars show polarization
values larger than 1\%.  
For the stars plotted in Fig. 1 a two-sided two-dimensional
Kolmogorov-Smirnoff test shows that the two samples (Vega-like
and normal field stars) are different to 99.7\%. \\

\begin{table*}
\caption[]{Estimated fractional infrared excesses $(L_{IR}/L_*)$  for Vega-like stars}
\scriptsize{
\begin{tabular}{llllll}\hline \hline

      HD     &  (IRAS  	  & Flux          &  Densities)       &             &      $L_{IR}/L_*$  \\ 
    Number   &       $F_{12}$ (Jy)  &    $F_{25}$ (Jy)     &      $F_{60}$ (Jy)	  &  $F_{100}$ (Jy)	  &          		\\ \hline \hline
      9672   & $ 3.31\times10^{-01}$ & $ 4.06\times10^{-01}$ &  $2.00\times10^{+00}$  & $1.91\times10^{+00}$ & $1.64\times10^{-3}$    \\
      17206  &	$1.88\times10^{+00}$ &  $6.54\times10^{-01}$ &  $1.63\times10^{+00}$  & $5.01\times10^{+00}$& 	$1.67\times10^{-3}$ \\
      17443  &	$4.19\times10^{-01}$ &  $2.39\times10^{+00}$ &  $1.45\times10^{+01}$  & $2.28\times10^{+01}$ &	$6.32\times10^{-2}$  \\
      32509  &	$4.33\times10^{-01}$ &  $6.87\times10^{-01}$ &  $2.48\times10^{+00}$  & $6.59\times10^{+00}$ &	$1.22\times10^{-2}$   \\
      34700  &	$6.05\times10^{-01}$ &  $4.42\times10^{+00}$ &  $1.41\times10^{+01}$  & $9.38\times10^{+00}$ &	$3.48\times10^{-1}$   \\
      37389  &	$3.84\times10^{-01}$ &  $1.40\times10^{+00}$ &  $1.33\times10^{+01}$  & $4.21\times10^{+01}$ &	$1.65\times10^{-1}$  \\
      53300  &	$7.48\times10^{-01}$ &  $3.15\times10^{-01}$ &  $4.14\times10^{-01}$  & $2.80\times10^{+00}$ &	$6.75\times10^{-3}$   \\
      93331  &	$2.50\times10^{-01}$ &  $2.54\times10^{-01}$ &  $6.71\times10^{-01}$  & $1.32\times10^{+00}$ &	$3.38\times10^{-3}$    \\
      98800  &	$1.98\times10^{+00}$ &  $9.28\times10^{+00}$ &  $7.28\times10^{+00}$  & $4.46\times10^{+00}$ &	$1.63\times10^{-1}$    \\
      99211  &	$1.57\times10^{+00}$ &  $4.27\times10^{-01}$ &  $4.00\times10^{-01}$  & $1.00\times10^{+00}$ &	$7.34\times10^{-4}$    \\
      102647 &	$6.97\times10^{+00}$ &  $2.11\times10^{+00}$ &  $1.18\times10^{+00}$  & $1.00\times10^{+00}$ &	$4.56\times10^{-4}$   \\
      109085 &	$2.18\times10^{+00}$ &  $7.73\times10^{-01}$ &  $5.44\times10^{-01}$  & $1.00\times10^{+00}$ &	$1.18\times10^{-3}$    \\
      115892 &	$3.56\times10^{+00}$ &  $9.72\times10^{-01}$ &  $4.00\times10^{-01}$  & $1.00\times10^{+00}$ &	$2.44\times10^{-4}$    \\
      121847 &	$2.39\times10^{-01}$ &  $3.62\times10^{-01}$ &  $1.08\times10^{+00}$  & $2.16\times10^{+00}$ &	$7.06\times10^{-4}$   \\
      123247 &	$1.33\times10^{-01}$ &  $1.56\times10^{-01}$ &  $5.13\times10^{-01}$  & $4.40\times10^{+00}$ &	$1.80\times10^{-3}$ \\
      131885 &	$1.06\times10^{-01}$ &  $2.14\times10^{-01}$ &  $3.52\times10^{-01}$  & $1.07\times10^{+00}$ &	$1.84\times10^{-3}$     \\
      135344 &	$1.59\times10^{+00}$ &  $6.71\times10^{+00}$ &  $2.56\times10^{+01}$  & $2.57\times10^{+01}$ &	$8.38\times10^{-2}$    \\
      139614 &	$4.11\times10^{+00}$ &  $1.81\times10^{+01}$ &  $1.93\times10^{+01}$  & $1.39\times10^{+01}$ &	$3.41\times10^{-1}$   \\
      139664 &	$1.42\times10^{+00}$  & $6.93\times10^{-01}$ &  $6.61\times10^{-01}$  & $2.37\times10^{+00}$ &	$1.44\times10^{-3}$    \\
      142096 &	$5.55\times10^{-01}$  & $3.89\times10^{-01}$ &  $6.46\times10^{-01}$  & $1.30\times10^{+00}$ &	$1.55\times10^{-4}$    \\
      142114 &	$6.64\times10^{-01}$  & $1.73\times10^{+00}$ &  $4.20\times10^{+00}$  & $1.16\times10^{+01}$ &	$4.02\times10^{-4}$    \\
      142165 &	$2.50\times10^{-01}$  & $3.40\times10^{-01}$ &  $2.77\times10^{+00}$  & $1.16\times10^{+01}$ &	$9.01\times10^{-4}$    \\
      142666 &	$8.57\times10^{+00}$  & $1.12\times10^{+01}$ &  $7.23\times10^{+00}$  & $5.46\times10^{+00}$ &	$2.32\times10^{-1}$    \\
      143006 &	$8.57\times10^{-01}$  & $3.16\times10^{+00}$ &  $6.57\times10^{+00}$  & $4.82\times10^{+00}$ &	$4.80\times10^{-1}$    \\
      145482 &	$6.58\times10^{-01}$  & $1.04\times10^{+00}$ &  $3.13\times10^{+00}$  & $1.11\times10^{+01}$ &	$3.78\times10^{-4}$    \\
      149914 &	$3.90\times10^{-01}$  & $5.25\times10^{-01}$ &  $3.53\times10^{+00}$  & $1.23\times10^{+01}$ &	$4.77\times10^{-3}$     \\
      233517 &	$5.02\times10^{-01}$  & $3.60\times10^{+00}$ &  $7.60\times10^{+00}$  & $5.10\times10^{+00}$ &	$9.62\times10^{-2}$    \\
      3003   & 	$4.78\times10^{-01}$  & $3.21\times10^{-01}$ &  $4.00\times10^{-01}$  & $1.00\times10^{+00}$ &	$8.26\times10^{-4}$    \\
      10476  &	$2.00\times10^{+00}$  & $6.64\times10^{-01}$ &  $4.00\times10^{-01}$  & $1.25\times10^{+00}$ &	$2.40\times10^{-3}$    \\
      10647  &	$8.22\times10^{-01}$  & $3.42\times10^{-01}$ &  $8.51\times10^{-01}$  & $1.08\times10^{+00}$ &	$2.10\times10^{-3}$    \\
      10700  &	$9.56\times10^{+00}$  & $2.16\times10^{+00}$ &  $5.12\times10^{-01}$  & $1.36\times10^{+00}$ &	$2.37\times10^{-3}$   \\
      14055  & 	$1.07\times10^{+00}$  & $4.70\times10^{-01}$ &  $8.58\times10^{-01}$  & $8.48\times10^{-01}$ &	$4.71\times10^{-4}$  \\
      16157  &	$5.43\times10^{-01}$  & $2.01\times10^{-01}$ &  $4.00\times10^{-01}$  & $1.07\times10^{+00}$ &	$1.35\times10^{-2}$    \\
      18978  & 	$1.38\times10^{+00}$  & $3.53\times10^{-01}$ &  $4.00\times10^{-01}$  & $1.00\times10^{+00}$ &	$6.75\times10^{-4}$    \\
      20010  &	$4.11\times10^{+00}$  & $9.70\times10^{-01}$ &  $2.87\times10^{-01}$  & $1.00\times10^{+00}$ &	$1.65\times10^{-3}$    \\
      22049  &	$9.52\times10^{+00}$  & $2.65\times10^{+00}$ &  $1.66\times10^{+00}$  & $1.89\times10^{+00}$ &	$2.46\times10^{-3}$   \\
      38393  &	$4.60\times10^{+00}$  & $1.08\times10^{+00}$ &  $4.00\times10^{-01}$  & $1.00\times10^{+00}$ &	$1.28\times10^{-3}$    \\
      43955  &	$1.47\times10^{-01}$  & $1.45\times10^{-01}$ &  $4.42\times10^{-01}$  & $1.46\times10^{+00}$ &	$1.00\times10^{-4}$   \\
      49336  &	$2.39\times10^{-01}$  & $1.74\times10^{-01}$ &  $1.72\times10^{-01}$  & $1.09\times10^{+00}$ &	$2.67\times10^{-4}$    \\
      53376  &	$6.86\times10^{-01}$  & $2.79\times10^{-01}$ &  $4.00\times10^{-01}$  & $4.46\times10^{+00}$ &	$6.85\times10^{-2}$    \\
      68456  &	$1.46\times10^{+00}$  & $2.72\times10^{-01}$ &  $4.00\times10^{-01}$  & $1.85\times10^{+00}$ &	$1.42\times10^{-3}$    \\
      69830  &	$9.66\times10^{-01}$  & $3.75\times10^{-01}$ &  $4.00\times10^{-01}$  & $1.00\times10^{+00}$ &	$2.80\times10^{-3}$    \\
      71155  & 	$1.15\times10^{+00}$  & $4.49\times10^{-01}$ &  $4.00\times10^{-01}$  & $1.00\times10^{+00}$ &	$4.13\times10^{-4}$     \\
      73390  &	$5.11\times10^{-01}$  & $2.50\times10^{-01}$ &  $5.32\times10^{-01}$  & $1.71\times10^{+00}$ &	$2.29\times10^{-4}$    \\
      95418  &	$4.80\times10^{+00}$  & $1.39\times10^{+00}$ &  $6.27\times10^{-01}$  & $1.00\times10^{+00}$ &	$3.85\times10^{-4}$    \\
      108483 &	$5.61\times10^{-01}$  & $2.64\times10^{-01}$ &  $4.00\times10^{-01}$  & $5.20\times10^{+00}$ &	$9.83\times10^{-5}$    \\
      114981 &  $1.56\times10^{-01}$  & $1.40\times10^{-01}$ &  $2.16\times10^{-01}$  & $7.38\times10^{-01}$ &	$1.00\times10^{-3}$   \\
      128167 & 	$1.60\times10^{+00}$  & $3.76\times10^{-01}$ &  $4.00\times10^{-01}$  & $1.00\times10^{+00}$ &	$1.12\times10^{-3}$    \\
      139006 & 	$5.92\times10^{+00}$  & $1.73\times10^{+00}$ &  $7.51\times10^{-01}$  & $1.00\times10^{+00}$ &	$3.76\times10^{-4}$    \\
      143018 &	$3.05\times10^{+00}$  & $5.47\times10^{+00}$ &  $9.25\times10^{+00}$  & $8.41\times10^{+00}$ &	$1.25\times10^{-4}$   \\
      149630 & 	$9.64\times10^{-01}$  & $2.47\times10^{-01}$ &  $2.64\times10^{-01}$  & $1.00\times10^{+00}$ &	$3.23\times10^{-4}$    \\
      153968 & 	$3.68\times10^{-01}$  & $3.77\times10^{-01}$ &  $4.00\times10^{-01}$  & $1.71\times10^{+00}$ &	$3.32\times10^{-2}$   \\
      161868 &  $1.41\times10^{+00}$  & $5.10\times10^{-01}$ &  $1.29\times10^{+00}$  & $2.35\times10^{+00}$ &	$4.14\times10^{-4}$   \\
      162917 & 	$5.44\times10^{-01}$  & $2.50\times10^{-01}$ &  $3.80\times10^{-01}$  & $1.26\times10^{+00}$ &	$1.60\times10^{-3}$   \\
      172167 & 	$4.16\times10^{+01}$  & $1.10\times10^{+01}$ &  $9.51\times10^{+00}$  & $7.76\times10^{+00}$ &	$3.55\times10^{-4}$   \\
      176638 &  $4.74\times10^{-01}$  & $3.95\times10^{-01}$ &  $4.00\times10^{-01}$  & $1.28\times10^{+00}$ &	$5.17\times10^{-4}$   \\
      181296 &  $5.39\times10^{-01}$  & $ 4.77\times10^{-01}$ &  $5.33\times10^{-01}$  & $1.00\times10^{+00}$ &	$7.63\times10^{-4}$   \\
      181869 & 	$8.21\times10^{-01}$  & $ 3.48\times10^{-01}$ &  $4.00\times10^{-01}$  & $1.00\times10^{+00}$ &	$3.16\times10^{-4}$   \\
      214953 & 	$6.26\times10^{-01}$  & $ 4.43\times10^{-01}$ &  $4.00\times10^{-01}$  & $1.00\times10^{+00}$ &	$3.16\times10^{-3}$  \\
      216956 &	$1.82\times10^{+01}$  & $ 4.81\times10^{+00}$ &  $9.02\times10^{+00}$  & $1.12\times10^{+01}$ &	$6.43\times10^{-4}$   \\
      224392 & 	$5.21\times10^{-01}$  & $ 1.93\times10^{-01}$ &  $4.00\times10^{-01}$  & $1.00\times10^{+00}$ &	$8.01\times10^{-4}$   \\  \hline

\end{tabular}
}
\end{table*}
\normalsize

    For the Vega-like stars that show relatively large degree of polarization,
the observed polarization can not be accounted for by normal interstellar
polarization and must be circumstellar in origin. Circumstellar dust
around these stars, to which the observed infrared excesses are ascribed, 
can cause polarization by the the process of scattering of the light from the
central star. In order to be able to produce a net polarization in the
integrated light the dust must be distributed in a non-spherical geometry.
These non-spherical distributions could be flattened disks around the
stars similar to those in $\beta $ Pic. The disk planes should have
relatively small inclinations with the observer's line of sight. Given
this constraint on the inclination of the disks, the difference between
the Vega-like stars and the normal stars noticed in Fig. 1 becomes even
more significant, because only a fraction of the stars with disks will have
have favourable inclinations. \\  
\begin{figure*}
\resizebox{\hsize}{!}{\includegraphics{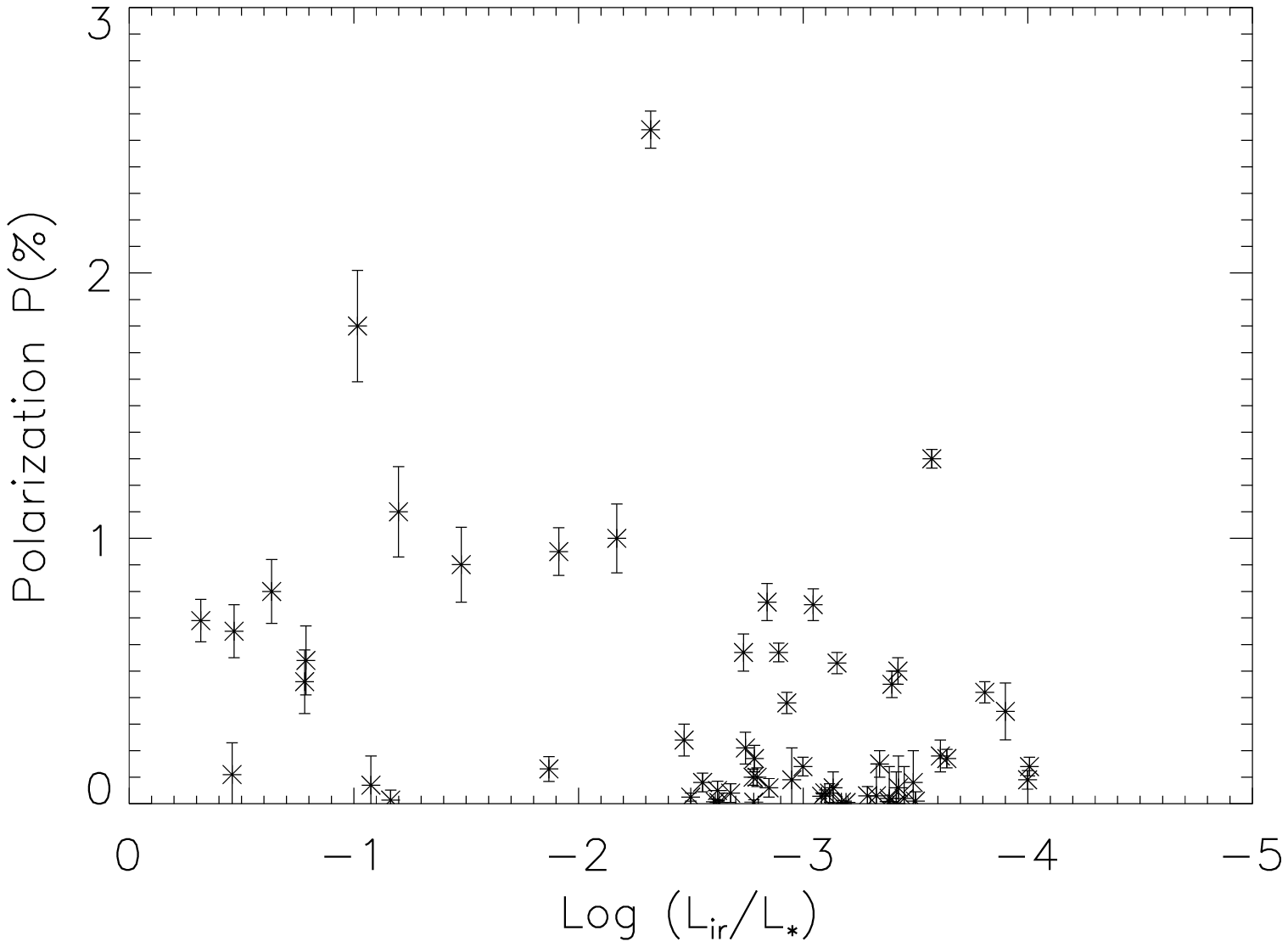}}
\caption{Plot of percentage polarization against the fractional infrared excess for Vega-like stars}
\label{Figure 4.}
\end{figure*}  

    The large values of polarization observed for Vega-like stars thus support 
the existence of dusty disks around these stars. The polarization is
produced by scattering of light from the central star. In contrast, the 
interstellar polarization is caused by selective extinction due to
nonspherical dust grains aligned by the intertellar magnetic field.
The interstellar polarization shows a correlation with reddening given by 
the average relation: $P_{V}/E(B - V) = 3 \% mag^{-1}$. The circumstellar
dust in Vega-like stars is believed to consist of relatively large grains
(a few tens of microns (Backman \& Parsce 1993)) that are not expected to produce any additional reddening at
optical wavelengths. Thus large values of polarization may result even
with small values of reddening at small distances. Fig. 3 shows a plot of
the observed polarization against reddening $E(B - V)$ for the Vega-like
stars, together with the line representing the average relation followed
by interstellar polarization and reddening. It can be seen that there are
more stars above this line than below it. Also, there is a lack of
correlation between polarization and reddening in Vega-like stars for
smaller values of reddening. Only at relatively large distances when the
interstellar component becomes large does the observed polarization
begin to generally increase with reddening. 

Circumstellar dust in Vega-like stars absorbs light from the central star 
and radiates thermally in the infrared. Dust also scatters starlight and
produces polarization. One may therefore expect some correlation between
the observed polarization in the optical and the excess infrared emission 
from these objects. In Fig. 4 we plot the observed percentage polarization
against the excess infrared luminosity ($L_{IR}$) as a fraction of the
total bolometric luminosity ($L_{*}$) of the star (the ratio $L_{IR}/L_{*}$)
for the Vega-like stars. The infrared excess $L_{IR}$ has been computed
from the observed flux densities in the IRAS bands  using  the 
equation (Cox 2000)  
$$ F_{IR}(7-135\mu m)=1.0 \times 10^{-14}(20.653f_{12} $$
$$+7.538f_{25}+4.578f_{60}+1.762f_{100}) Wm^{-2} $$
for the total infrared flux and the appropriate distance factor 
$4\pi d^2$ where d is the distance to the star.
The stellar bolometric luminosity $L_{*}$ corresponds to the
bolometric luminosity of a normal star of the same spectral type as that
of the Vega-like star under consideration. The $ L_{IR}/L_{*}$  values are listed in Table 3.
A positive correlation is
apparent from Fig. 4. Vega-like stars with relatively larger infrared
excesses tend to have larger values of polarization in the optical. The
ratio $L_{IR}/L_{*}$ can be taken to be a rough measure of the dust
optical depth $\tau$ in the optical where the absorption of starlight takes
place. The observed values of the ratio $L_{IR}/L_{*}$ imply $\tau$ in the
range $\sim 10^{-4}$ - $\sim 10^{-1}$. For large dust grains (grain size $a$
$>>$ $\lambda$, the wavelength of light), that are generally thought to
dominate the dust disks of Vega-like stars, the absorption and scattering
efficiencies can be taken to be $\sim 1$. If the polarization is produced
by scattering of starlight by the same dust that absorbs stellar radiation
and emits thermally in the infrared, then for the observed range of the
$L_{IR}/L_{*}$ ratio the expected range of polarization values would be    
$\sim 10^{-2} \% - \sim 10 \% $. Stars with $L_{IR}/L_{*} \sim 10^{-2}$
would have polarization up to the level of $\sim 1\%$. Here it is
assumed that the circumstellar material is optically thin so that
there is single scattering only (Bastien 1987). The maximum
linear polarization that can be produced in ellipsoidal models
with single scattering is about 1.1\% (Shawl 1975). Larger values
of polarization can result if the direct light from the central
star is extincted by obscuring dust in front of the star. Also, the
observed polarization depends on the flatness and orientation of
the dust disk.
It is to be noted from Fig. 4 that several
stars with $L_{IR}/L_{*}$ $\sim$ $10^{-3}$ or less also show polarization
$\sim 0.5 \%$, much larger than the expected values $< \sim 0.1 \%$. It is
possible that these stars have an additional dust component with smaller
grains with high albedo. This population of grains would cause
relatively large polarization by scattering without producing large
infrared excesses.

\section{Conclusions}
The results of the present study of the polarimetric behaviour of
Vega-like stars can be summarized as follows.
\begin{itemize}
\item  Vega-like stars generally show optical linear polarization that
is much larger than that which can be ascribed to interstellar polarization
along the line of sight to these relatively nearby objects.
\item The anomalous polarization in Vega-like stars is caused
by scattering of stellar light due to circumstellar dust 
 distributed in non-spherically symmetric envelopes, and is
fully consistent with a distribution in flattened disks.
\item The absence of any excess reddening in these stars is consistent
with the dust grains in their circumstellar disks being relatively large 
in size.
\item In some Vega-like stars that show relatively small infrared excesses
but large values of polarization, an additional component of dust
consisting of smaller grains with high albedo may also be present.
\end{itemize}
Multiwavelength polarization measurements of Vega-like stars would be able
to shed more light on the nature and distribution of the dust in their
circumstellar environment.  \\

\begin{acknowledgements}
                        We would like to thank the referee Prof.
P. F. Bastien for several very useful comments.
\end{acknowledgements}

\end{document}